**Unbreakable distributed storage with quantum key distribution network and password-authenticated secret sharing**

M. Fujiwara[1*], A. Waseda[2], R. Nojima[2], S. Moriai[2], W. Ogata[3] & M. Sasaki[1]

[1]Quantum ICT Laboratory, National Institute of Information and Communications Technology (NICT), 4-2-1 Nukui-kita, Koganei, Tokyo 184-8795, Japan

[2]Security Fundamentals Laboratory, National Institute of Information and Communications Technology (NICT), 4-2-1 Nukui-kita, Koganei, Tokyo 184-8795, Japan

[3]Tokyo Institute of Technology, 2-12-1 Ookayama, Meguro-ku, Tokyo 152-8552 Japan

*fujwara@nict.go.jp

**Abstract**

Distributed storage plays an essential role in realizing robust and secure data storage in a network over long periods of time. A distributed storage system consists of a data owner machine, multiple storage servers and channels to link them. In such a system, secret sharing scheme is widely adopted, in which secret data are split into multiple pieces and stored in each server. To reconstruct them, the data owner should gather plural pieces. Shamir's ($k, n$)-threshold scheme, in which the data are split into $n$ pieces (shares) for storage and at least $k$ pieces of them must be gathered for reconstruction, furnishes information theoretic security, that is, even if attackers could collect shares of less than the threshold $k$, they cannot get any information about the data, even with unlimited computing power. Behind this scenario, however, assumed is that data transmission and authentication must be perfectly secure, which is not trivial in practice. Here we propose a totally information theoretically secure distributed storage system based on a user-friendly single-password-authenticated secret sharing scheme and secure transmission using quantum key distribution, and demonstrate it in the Tokyo metropolitan area (≤90km).

**Introduction**

Nowadays highly confidential data such as personal genetic information are stored in data centers and storage area networks. In these systems, information leakage to a system provider and attackers on the storage servers is the most likely risk, and when happened, it causes serious damage to data owners. Security of such data should be tightly protected not only in storage for a long period of time but also in data transmission between the storage servers, which should often be located in distant places for robust site diversity. Confidentiality of data in storage is usually guaranteed by encrypting them. Some recent cryptographic schemes enable a data owner to search over encrypted data or process data without decryption [1,2]. Especially, lattice-based cryptography [3-6] attracts attention due to its resistance against quantum algorithms[7] and provable security under some worst case hardness assumptions. However, evaluation on the security of lattice-based cryptography is an ongoing task. For example, NTRU lattice cryptography which has been standardized at IEEE in 2009[8] is threatened by the newly developed efficient attack algorithm[9] and needs to be re-evaluated on the security level.

In contrast, Shamir's secret sharing (SS) scheme[10] can realize information theoretically secure storage systems, if data-transmission and authentication in the systems could be performed somehow in an information theoretically secure way. Some protocols based on SS further allow processing of shared data without reconstruction[11]. Practical implementations of data-transmission and user authentication with information theoretical security (ITS) are, however, not easy at all. The only the known way for data-transmission with ITS is to use one-time pad (OTP) with a truly random number key stream, which can be shared either by trusted couriers or quantum key distribution (QKD)[12]. As for message and user authentication, the Wegman-Carter scheme[13] can be used to ensure ITS, but the data owner has to employ pre-shared keys for each storage server, and spend them in the OTP manner. It requires frequent key sharing and rigorous key management. So in practice, most data storage services adopt an easier way, i.e., password authentication for on-line individual identifications due to its high usability. If the data owner registers the same password to all storage servers, attackers who can access at least one storage server is able to easily know the password, and then access all the storage servers. Even if only hashed password is stored, a powerful malicious insider who can access a password file in a storage server may guess the password with an off-line dictionary attack. That is, by hashing all possible passwords and comparing with the registered hash value, the malicious insider can find out the correct password without making authentication transaction. To attain ITS against the malicious insider's off-line dictionary attacks, different passwords should be used for different storage servers at each time. So the data owner has to remember many passwords, and hence often tends to reuse a password or employ easy passwords, which introduces the vulnerability against an on-line dictionary attack even if the number of attack trials is limited. Although password-authenticated SS schemes[14,15] based on homomorphic encryption use a single password and have high tolerance even against off-line dictionary attacks, they so far offer only computational security[16].

Since critical data which should be securely stored for long periods of time are rapidly increasing, demands for an unbreakable distributed storage system ensured by ITS are increasing. If information theoretically secure password-authentication protocol could be implemented, it would be beneficial to SS schemes. To our best knowledge, however, such a storage system has never been proposed and demonstrated so far.

Here we newly develop and demonstrate an efficient information theoretically secure distributed storage system by combining quantum key distribution and password-authenticated secret sharing. In our system, ITS in user authentication can be realized by embedding a single-password authentication mechanism into the reconstruction process of an SS scheme, while ITS message authentication is ensured by the Wegman-Carter scheme. ITS in data transmission can be realized by using OTP with random number key streams supplied from "the QKD platform[17]". The whole system operates by the interplay between the QKD platform and the application layer on it, in which password-authenticated secret sharing is implemented. Moreover, we show how the QKD platform is designed in a layer structure for high serviceability and availability of the secure storage system.

**Method**

**Information-theoretically secure password-authenticated SS scheme**

Our newly proposed password-authenticated secret sharing scheme is based on Shamir's SS scheme. In the original Shamir's SS scheme, secret data $D$ is divided into $n$ pieces of shares $f_D(a_1), f_D(a_2), \ldots, f_D(a_n)$, where $f_D$ is a random polynomial of degree at most $k-1$ with a free coefficient representing the secret data $D$ itself, and $a_1, a_2, \ldots, a_n$ are public values. Then, knowledge of any $k$ or more pieces of $f_D(a_i)$ makes $D$ easily computable with Lagrange interpolation. For example, to determine the free coefficient $D$ in the quadratic function $f_D(x)$, at least three shares are required because there are three unknown parameters in the quadratic function $f_D(x)$. On the other hand, knowledge of any $k-1$ or fewer pieces of $f_D(a_i)$ leaves $D$ completely undetermined (in the sense that all its possible values are equally likely). That is, the attacker cannot recover the original data from less than the threshold $k$ of shares even by using unlimited computational resources. Such a scheme is called ($k, n$)-threshold scheme. This enables secret calculations (addition and multiplication). In fact $f_{D^{(1)}}(a_i) + f_{D^{(2)}}(a_i)$ becomes a share of addition of two secret data $D^{(1)}$ and $D^{(2)}$. Likewise, $f_{D^{(1)}}(a_i) \times f_{D^{(2)}}(a_i)$ is used as the share of $D^{(1)} \times D^{(2)}$. In multiplying process, however, the degree of the polynomial $f_{D^{(1)}}(x) \times f_{D^{(2)}}(x)$ is $2k-2$. So, $2k-1$ of shares are necessary to reconstruct $D^{(1)} \times D^{(2)}$.

Now we describe our password-authenticated secret sharing scheme. It comprises three phases: (1) **Registration phase** in which data shares and password shares are computed and stored in the storage servers; (2) **Pre-computation and communication phase** in which each storage server generates a random number by a physical random number generator, and computes their shares, as well as prepares shares of the data "0" with another random number for confidentiality of the share in data reconstruction without changing the value of data share. Then all the storage servers send the shares to each other; (3) **Data reconstruction phase** in which the data owner requests the storage servers to send back the data shares using the password (request-response). If the password is correct, then the original data is reconstructed.

In these phases, all communications between the data owner machine and the storage servers and among the storage servers are to be OTP-encrypted by the keys supplied from the QKD platform. Thus

ITS in data transmission can be ensured.

As for ITS in user authentication, the mechanism is as follows. The required task is to prevent an attacker from collecting knowledge nothing more than obtained by the on-line dictionary attack. In our scheme, this is realized by randomizing the responses from the storage servers by masking the data shares with the random numbers generated in **Pre-computation and communication phase** which should be discarded at each request-response. Only the correct password can cancel out this masking, and any wrong password cannot unmask the secret data. ITS in message authentication is ensured by the Wegman-Carter scheme.

A detailed procedure is exemplified below in the case where there are four storage servers ($n = 4$) denoted as 1, 2, 3, and 4, and we assume that an attacker can corrupt at most one storage server. In this case, the secret data $D$ are shared by the (3,4)-threshold scheme. On the other hand, the password shares must be generated by using the (2,4)-threshold scheme due to the multiplying process. This case suffices the demonstration of the principle of the information theoretically secure distributed storage. The schematic is shown in Fig. 1. All calculations are made in a finite Galois field with prime order $q$. The $q$ should be large enough for security, since the number of possible passwords is $q$ at most. We take $q$ from Mersenne primes, because they have a suitable form $q = 2^m - 1$ for calculations in the finite field. In our case, $m$ is chosen in a range of $10^3$-$10^5$.

(1) Registration phase

(1-1) Since each calculation in the finite field with prime order $q = 2^m - 1$ can deal with only blocks of length at most $m - 1$ bits, secret data $D$, which has generally a much longer length, needs to be divided into pieces of $(m - 1)$-bit block, say $l$ pieces; $D = D_l|D_{l-1}|\cdots|D_1$. The data owner sets a $(m - 1)$-bit password $P$, which should have sufficient entropy against the on-line dictionary attack, then computes a message authentication code, $MAC = D_l P^l + D_{l-1}P^{l-1} + \cdots + D_1 P$, which is denoted as $D_{l+1}$, and finally adds it to the data for later purpose of message authentication.

(1-2) For each data block, data shares $f_{D_i}(1)$, $f_{D_i}(2)$, $f_{D_i}(3)$, $f_{D_i}(4)$ are created for storage server 1, 2, 3, and 4, respectively, by using polynomial $f_{D_i}$ of degree at most 2, where $i = 1, \ldots, l + 1$. Password shares $f_P(1)$, $f_P(2)$, $f_P(3)$, $f_P(4)$ are created by using polynomial $f_P$ of degree at most 1.

(1-3) They are then sent to the corresponding storage servers.

(1-4) Each server stores the set of shares.

(2) Pre-computation and communication phase

(2-1) Each server generates a random number, denoted as $R_j$ for the $j$-th storage server, and makes its shares $f_{R_j}(1), f_{R_j}(2), f_{R_j}(3), f_{R_j}(4)$ by using polynomial $f_{R_j}$ of degree at most 1.

Furthermore each server generates shares of the "0" $f_{0j}(1), f_{0j}(2), f_{0j}(3), f_{0j}(4)$ by using polynomial $f_{0j}$ of degree at most 2, such that $f_{0j}(0) = 0$ should hold so as to keep confidentiality of the share in data reconstruction phase without changing the value of the data share.

(2-2) The storage servers send these shares to each other.

(2-3) Each server receives three shares of three random numbers and three shares of the "0," and stores them together with the ones produced by itself.

For ITS, the above procedure has to be iterated $l+1$ times before each data reconstruction of $l$ blocks secret data. That is, $j$-th storage server has to keep $l+1$ sets of $(f_{R1}(j), f_{R2}(j), f_{R3}(j), f_{R4}(j), f_{01}(j), f_{02}(j), f_{03}(j), f_{04}(j))$.

(3) Data reconstruction phase

Let $P'$ be the password in the data owner's memory.

(3-1) The data owner chooses three storage servers among the four. We may assume that they are storage server 1, 2, and 3 without loss of generality, denoting them as a set $L=\{1, 2, 3\}$.

(3-2) The data owner generates shares of $P'$, $f_{P'}(1)$, $f_{P'}(2)$, $f_{P'}(3)$ by using polynomial $f_{P'}$ of degree at most 1.

(3-3) Each set $(L, f_{P'}(j))$ is sent to each corresponding storage server (request).

(3-4) If $|L| \neq 3$, the request is rejected regarding it as an improper request. Otherwise, for each data block, each server, say $j$-th one, computes $R = f_{R1}(j) + f_{R2}(j) + f_{R3}(j)$, $Z = f_{01}(j) + f_{02}(j) + f_{03}(j)$ and
$$F_{ji} = (f_P(j) - f_{P'}(j))R + Z + f_{D_i}(j).$$
The $F_{ji}$ $(i = 1, \ldots, l+1)$ are then sent to the data owner (response). Here note that $R$ and $Z$ should be discarded at each request-response for ITS.

(3-5) For each data block, the data owner finds polynomial $F_i(x)$ of degree 2 that satisfies $F_i(j) = F_{ji}$ for all $j$. $F_i(0)$ is the reconstructed block.

(3-6) The data owner calculates $MAC$ from $F_1(0), \ldots, F_l(0)$ as in the first phase. If $F_{l+1}(0)$ is equal to calculated $MAC$, the data owner successfully reconstructs the secret data $D$.

In general, the number of servers $n$ and the (expected) maximum number of corrupted servers $t$, can be set arbitrarily provided as $n \geq 2t + 1$ is met. In such a general case, the degree of polynomial $f_P$, $f_{Rj}$, and $f_{P'}$ are $t$, while that of $f_{D_i}$ and $f_{0j}$ are $2t$, and the number of chosen servers $|L|$ is $2t + 1$. The number of servers should be set considering with cost and the risk of information leakage in each storage server.

In step (3-5), $F_i(x) = (f_P(x) - f_{P'}(x))f_R(x) + f_0(x) + f_{D_i}(x)$ holds. Therefore, if $P' = P$ (namely $f_P(0) = f_{P'}(0)$), $F_i(0) = D_i$ holds. In this way, by using responses from $2t + 1$ of $n$ servers, the data owner decodes the secret data $D$. On the other hand, the responses do never leak any information about $D$ if $P' \neq P$, since they are masked by $f_{Rj}$ and $f_{0j}$. More precisely, we can prove the following three facts; (i) if $t$ corrupted storage servers jointly try to forge the reconstructed data by active attacks on the protocol, the data owner can detect it with probability $(1 - l/q)$ if $P$ is chosen randomly, or equivalently the *MAC* can be forged with the probability of $l/q$, (ii) the total information which $t$ corrupted storage servers can see in the protocol is independent from the data owner's password $P$ and stored secret data $D$, if the other servers and the data owner are honest, (iii) even if an attacker first corrupts $t$ of storage servers, then participates in Data reconstruction phase pretending to be a data owner by utilizing the corrupted servers, the total information which the attacker can obtain is no information other than whether the guessed password $P'$ is equal to the correct password $P$ or not. Note that a guessed password $P'$ can be uniquely determined from values received by honest storage servers, even if the attacker pretends to be a data owner and sends random numbers in step (3-3) instead of $f_{P'}(j)$ computed from a certain $P'$. Similar to the normal SS scheme, if the number of corrupted servers exceeds $t$, the confidentiality of the secret data cannot be guaranteed.(See Supplemental information for the detail.)

**Figure 1. Schematic diagram of distributed storage with quantum key distribution and password-authenticated secret sharing scheme.**

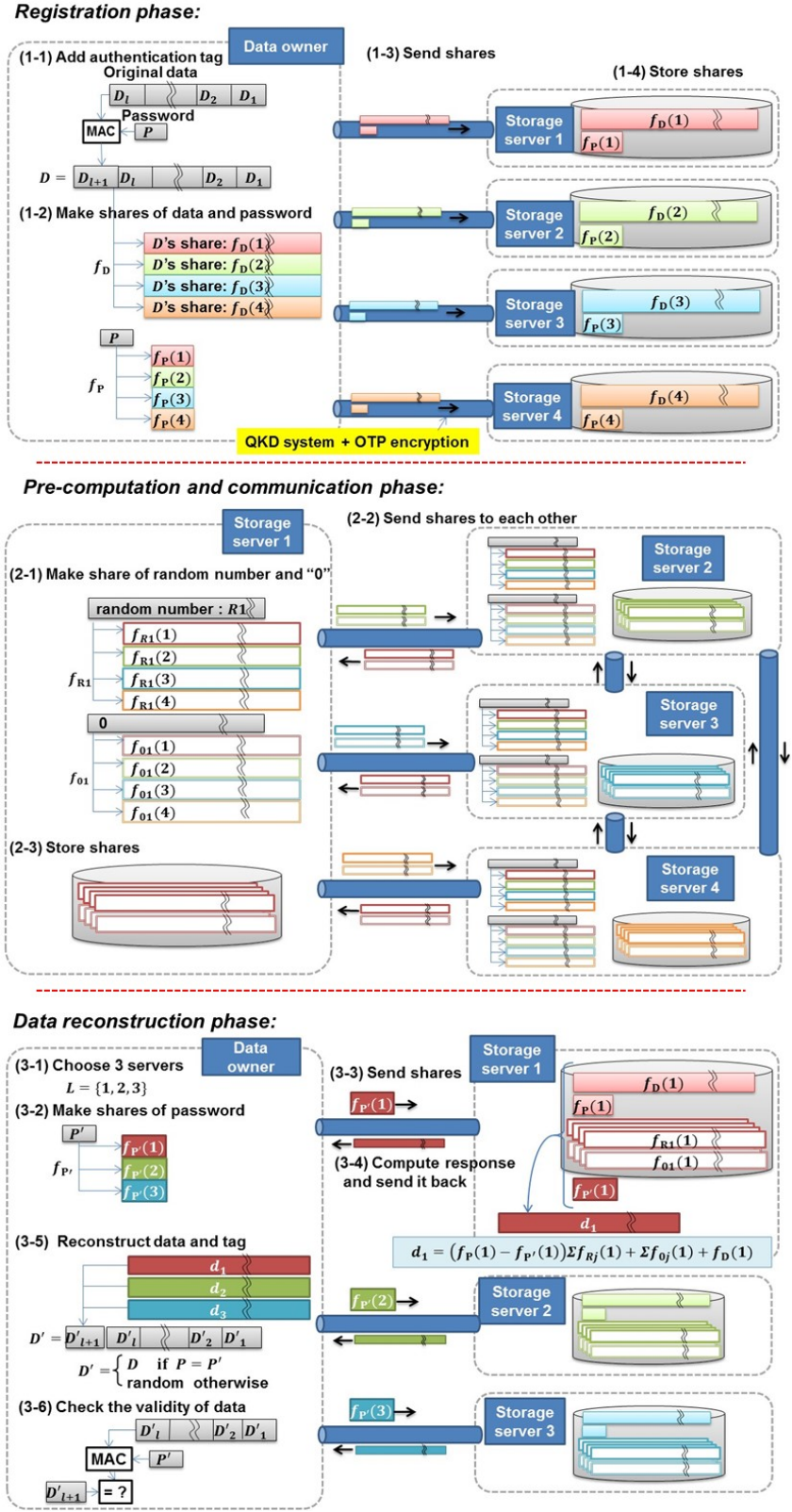

**Structure of the whole system**

To realize the scheme described above in a practical form, we adopt a layer architecture consisting of two blocks as shown in Fig.2, namely the application layer on which the password-authenticated secret sharing scheme is implemented and the QKD platform[17] which works as a secure key supply infrastructure. The data owner and the storage servers request random number keys to the QKD platform in every appropriate phase, and get them in an appropriate file format of key streams.

The QKD platform further consists of two layers, a quantum layer and a key management layer. In the quantum layer, each point-to-point QKD link generates symmetric random number key pairs. QKD link consists of transmitter and receiver devices, and two channels connecting them; one is a dedicated fiber for transmitting quantum signals, and the other is a public authenticated channel for key distillation processing.

Generated keys are pushed up to servers, called key management agents (KMAs), in the key management layer, then stored and managed. Each KMA must be located in a physically protected place, referred to as "trusted key relay node". The KMAs are connected by authenticated channels, and execute key relays by key encapsulation in a hop-by-hop fashion. Thus the key pair can be shared between two terminal nodes even if they are not directly connected by a QKD link. The KMA resizes the key strings into appropriate key files, saves them with key IDs and QKD transmitter/receiver IDs, and authenticates the relayed key file and other KMAs in the relay route.

In the key management layer, a key management server (KMS) is also located at one of the trusted key relay nodes, and gathers link information, including bit error rates, key rates, amounts of accumulated keys, from the KMAs, and organizes a routing table, and provisions secure paths to the KMAs. To make the interface with the application layer, a key supply agent (KSA) is introduced on top of each KMA, and supplies a user (the data owner/the storage server) the keys in an appropriate format depending on applications, according to the request from the user. The KSA records key IDs, users' and applications' IDs, and date of key usage, which are also sent to the KMS to guarantee the traceability of the key. When stored keys are expired, they are erased. Thus a key life cycles is properly managed, and the whole of the QKD platform is centrally controlled by the KMS.

The data owner and the storage servers can then use information theoretically secure keys in each phase. Once supplied with the keys, the responsibility of key management belongs to application users themselves. Thus the boundary of responsibility is set between the QKD platform and the application layer. Of course, access rights to the QKD platform and storage servers are completely separated.

**Figure 2. Schematic view of the layer structure of our distributed storage system.**

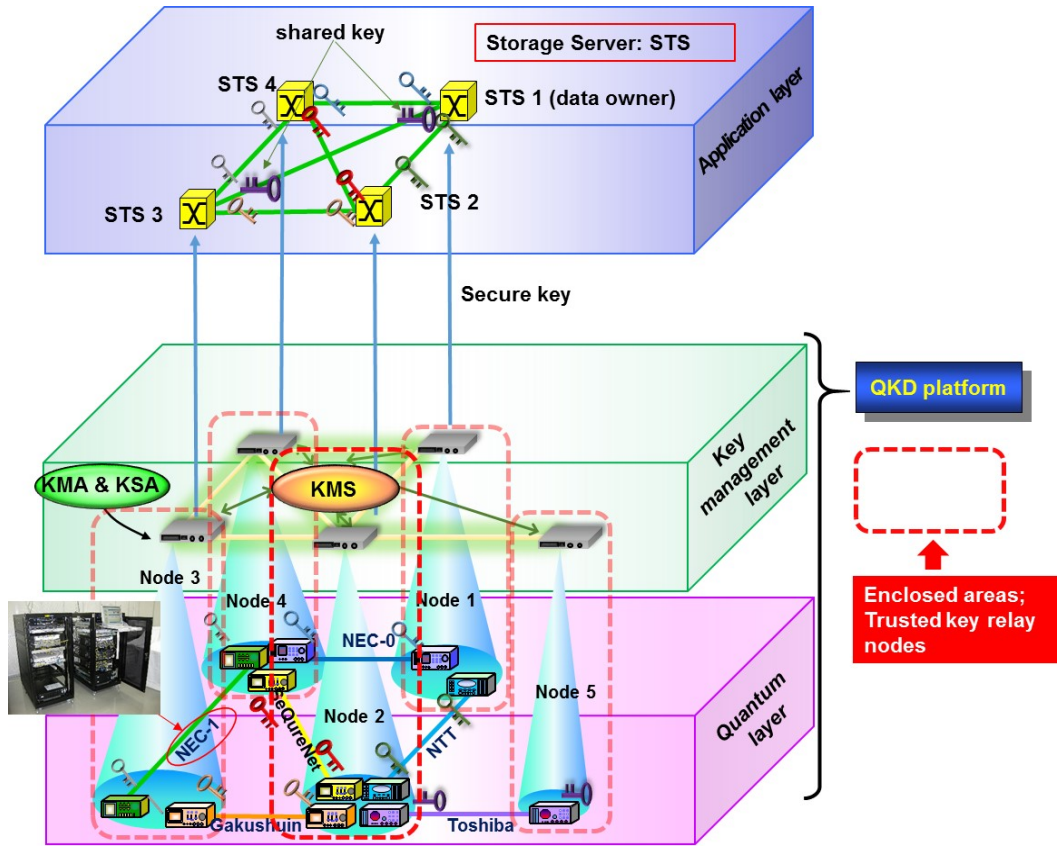

The whole system consists of two blocks, the application layer and the QKD platform. The QKD platform further consists of the two layers, the quantum layer and the key management layer, working as a secure key supply infrastructure. The password-authenticated secret sharing scheme is implemented in the application layer on which the data owner and the storage servers (STSs) are setup. In the QKD platform, the keys are generated in each QKD link in the quantum layer, pushed up to key management agents (KMAs) in the key management layer. The KMAs are in the trusted key relay nodes, store the keys, and if necessary, relay the keys. To support various applications, key supply agents (KSAs) are introduced at each KMA. In the key management layer, a key management server (KMS) is also located and carry out the centralized key management. Having requests from the data owner and the storage servers, the KSAs supply them the random number key streams.

## Results

To test practicality of our system, we measure throughputs of the system on the QKD platform in the field environment. We have operated the QKD platform, constructed in a field testbed covering the Tokyo metropolitan area (≤90km), named as the Tokyo QKD network. The Tokyo QKD Network itself consists of five nodes connected by six QKD links. As inscribed in Fig. 2, NEC Corporation provides two QKD links[18]. In these links, BB84[19] protocol with decoy states[20,21] are used. The QKD link "NEC-1" is made by a 22 km long field installed fiber looped back with the total attenuation rate of 13dB. About 95% of this fiber is aerial over poles. The Toshiba QKD link is based on decoy state BB84 protocol, and connects the NICT headquarter and Otemachi (very center of Tokyo) through a dark fiber of the so-called JGN-X testbed[22]. Its length is 45 km and the transmission loss is about 14.5dB. The Gakushuin QKD link uses the Continuous Variable (CV)-QKD[23]. It is deployed in a link of about 2 km in the NICT premise. Another CV-QKD link provided by SeQureNet SARL is also deployed in the NICT premises[24]. The NTT QKD link is based on Differential Phase Shift (DPS)-QKD, in a loop-back configuration between the NICT headquarter and Otemachi whose length is 90km with an attenuation rate of 28.6dB[25]. Superconducting Single Photon Detectors (SSPDs)[26] are used in this link due to their low dark counts. The quantum efficiencies of SSPDs are 10% at 100 c/s of dark count rate. Specifications are listed in table 1.

The information theoretically secure distributed storage system with (3,4) threshold for secret data is physically constructed with four nodes which are denoted as Node 1 to 4 of the QKD platform in Fig. 2, and the data owner is located in the same node as "storage server 1" just for experimental convenience. Here we should assume that in this node, "storage server 1" may be corrupted by attackers but the data owner can be protected against attackers. All the communications between the data owner machine and the storage servers and among the storage servers are carried out based on the QKD-enhanced Internet protocol (IP)[27], in which all IP packets are OTP-encrypted and further authenticated with *MAC* based on Wegman-Carter protocol by using the keys from the QKD platform. The maximum payload size of IP frame format is 1500 bytes.

We prepare three kinds of secret data $D$ whose size is 6955, 13695, and 46000 bytes. We prepare Mersenne primes $q = 2^m - 1$ with indices $m$=521, 1279, 2203, 3217, 4253, 10041, 11213, 19937, 23209, 44497, and 86243. The experimental results are shown in Fig. 3. These results show that our system achieves practical processing times. As shown in Fig. 3, performance of our system depends on the size of $q$. This is because (1) the computational time of the shares increases roughly in the square of bit length of $q$ and (2) using a smaller prime $q$ increases the number of blocks $l$, and hence a longer processing time is required for dividing/managing the blocks and sending IP packets. There would be a good balance point, conditioned on the maximum payload size of 1500 bytes. In our experiments, the best performance can be found in the range of $q$ with $11213 \leq m \leq 23209$. The processing time of OTP-encryption of IP packets and calculation of *MAC* is negligible compared with calculation time of the other processes. The time of total process would be improved by implementing a plurality of payloads into a single frame with considering of the maximum payload size of 1500 bytes. ITS in our distributed storage is more or less realized at the expense of the amount of keys from the QKD platform. Actually, the total length of keys required to store and retrieve is about 30 times as long as the original secret data. Anyway, we show that the QKD platform works as the stable secure key supply infrastructure for requests from multi-users.

**Table 1. Specification of QKD links**

|  | Protocol | Transmission | |
|---|---|---|---|
|  |  | Length (km) | Loss (dB) |
| NEC-0 | BB84 with decoy | 50 (Spooled fiber NICT premise) | 10 |
| NEC-1 | BB84 with decoy | 22 (field installed 95% areal line) | 13 |
| Toshiba | BB84 with decoy | 45 (field installed 50% areal line) | 14.5 |
| NTT-NICT | DPS-QKD | 90 (field installed 50% areal line) | 28.6 |
| Gakushuin | CV-QKD | 2 (NICT premise) | 2 |
| SeQureNet | CV-QKD | 2 (NICT premise) | 2 |

**Figure 3. Processing time as a function of index of Mersenne prime for each phase.**

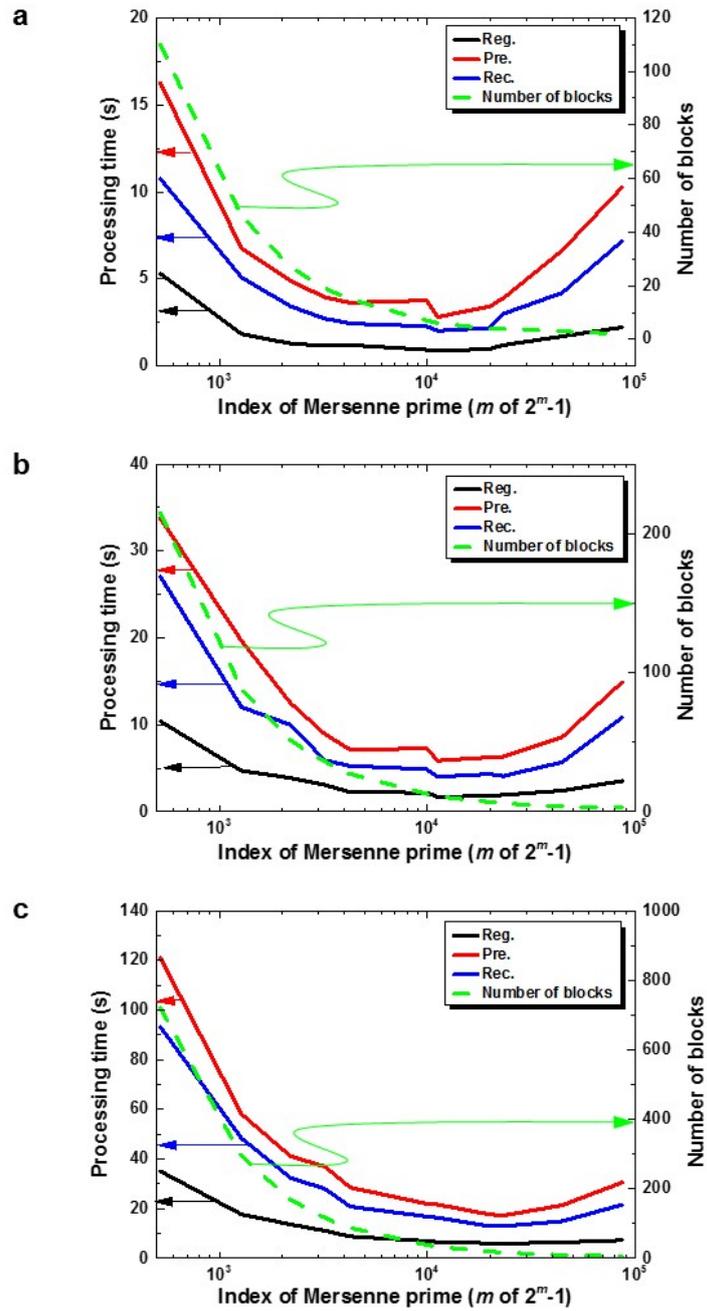

Reg : Registration phase (black). Pre : Pre-computation and communications phase (red). Rec : Data reconstruction phase (blue). Dependence of numbers of divided blocks on Mersenne prime size is shown by green dashed line**.** The sizes of secret data are, **a**, 6955 bytes, **b**, 13695 bytes, and **c**, 46000 bytes respectively.

## Discussion

Our newly proposed unbreakable distributed storage system will also be useful to realize secure data relay via classical nodes. And, depending on a QKD network topology, we can relax requirements to the current trusted key relay node in the QKD network, not resorting to a greatly costly quantum repeater paradigm. In the newly proposed data relay scheme, the storage servers are in the distributed relay nodes between the sender (Alice) and the receiver (Bob). Alice sends shares of her data to the distributed relay nodes, and Bob reconstructs the data via these relay nodes based on our distributed storage scheme. If relay nodes more than the threshold were never corrupted, and the other relay nodes could be trusted, then the distributed key relay with ITS could be realized in the sense of the ($k$, $n$)-threshold SS. Moreover, data can be securely stored in the distributed relay nodes, and can be downloaded when needed. If Alice and Bob can establish independent QKD links to each distributed relay node (embedded in the key relay node), this new data relay scheme could greatly relax the assumptions on the key relay nodes in the QKD network, and hence could be a practical option in QKD networks which consist of a relatively large number of key relay nodes.

## Conclusion

Thus by combining QKD and newly developed password-authenticated secret sharing scheme, we demonstrated, for the first time to our best knowledge, a distributed storage system with information theoretically secure data transmission, storage, and authentication in a metropolitan area network. This system can also implement secure data relay function. In this newly developed protocol, the throughputs enhancement would be necessary. However, such a development issues are made explicit by operation in the real QKD network for the first time. Our scheme and system would contribute greatly to dealing with highly confidential data, such as state secrets and personal genomic data, in a network over long periods of time, furnishing the site diversity as well.

## Acknowledgements

The authors thank all participants in the NICT Commissioned Research on Research and Development on Secure Photonic Network Technology for the collaboration to operate the Tokyo QKD Network. This work was partly funded by ImPACT Program of Council for Science, Technology and Innovation (Cabinet Office, Government of Japan).

## Author information

**Affiliations**

**Quantum ICT Laboratory National Institute of Information and Communications Technology (NICT), 4-2-1 Nukui-kita, Koganei, Tokyo 184-8795, Japan**

Mikio Fujiwara, Masahide Sasaki

**Security Fundamentals Laboratory National Institute of Information and Communications Technology (NICT), 4-2-1 Nukui-kita, Koganei, Tokyo 184-8795, Japan**

Atsushi Waseda, Ryo Nojima, Shiho Moriai

**Tokyo Institute of Technology, 2-12-1 Ookayama, Meguro-ku, Tokyo 152-8552 Japan**

Wakaha Ogata

**Contributions**

M. F. and M. S. conceived the design of information theoretically secure storage system based on QKD, and implemented it in the Tokyo QKD Network. A. W, R. N., S. M., and W. O. proposed the password-authenticated secret sharing scheme with the security proof. M. F. and M. S. wrote the manuscript with inputs from all authors.

**Competing interest**

The authors declare no competing financial interests.

**Corresponding author**

Correspondence to Mikio Fujiwara

**Supplementary information**

**Manuscript title:**

**Unbreakable distributed storage with quantum key distribution network and password-authenticated secret sharing**

Author list:

M. Fujiwara, A. Waseda, R. Nojima, S. Moriai, W. Ogata & M. Sasaki

**Sketch of security proof of password-authenticated secret sharing**

The scheme has the following three properties.

**Theorem 1**   If $t$ corrupted storage servers try together to change the reconstructed data by deviating from the protocol, the data owner can detect it with probability of $(1 - l/q)$ if password $P$ is randomly chosen.

(Proof)   Manipulating 0's shares sent to honest servers in **Pre-computation and communication phase** and the responses sent back to the data owner in **Data Reconstruction phase,** $t$ corrupted storage servers make the data owner reconstruct forged data blocks $\tilde{F}_i \neq F_i(0)$. The probability that *MAC* of forged data blocks $(\tilde{F}_l, ..., \tilde{F}_1)$ equals to $\tilde{F}_{l+1}$ is $l/q$, because *MAC* is calculated by using an $l$-degree polynomial that has (at most) $l$ solutions. This means that such malicious behavior can be detected with probability of $(1 - l/q)$.

**Theorem 2**   The total information which $t$ corrupted storage servers can see in the protocol is independent from the data owner's password $p$ and stored secret data $D$, if the other servers and the data owner follow the protocol.

(Proof)   We consider an attacker who corrupts $t$ servers. The information that the attacker can see in **Registration phase** is $t$ shares of password $P$ and secret data $D$, which are generated by using $t$-degree polynomials. Therefore, they are independent from $P$ and $D$ as in Shamir's SS. The information that the attacker can see in **Pre-computation and communication phase** is $t$ shares of random numbers and "0" computed by honest servers, which are clearly independent from $P$ and $D$.

The information that the attacker can see in **Data reconstruction phase** is $t$ shares of $P'$. Even if $P' = P$, the shares are independent from $P$ and $D$.

Note that this situation does not change even if the corrupted servers send fake shares to honest servers.

**Theorem 3**   Even if an attacker first corrupts $t$ of storage servers, then participates in Data reconstruction phase pretending to be a data owner by utilizing the corrupted servers, the total information which the attacker can obtain is no information other than whether the guessed password $P'$ is equal to the correct password $P$ or not.

(Proof)   Here we describe the proof in the case where there is only one data block. The generalization to the case of multiple data blocks is straightforward.

We consider an attacker who corrupts at most $t$ servers $C = \{c_1, \ldots, c_t\} \subseteq \{1, \ldots, n\}$ and tries to get some information on password $P$ and/or secret data $D$ by pretending the data owner in **Data reconstruction phase**. In **Data reconstruction phase**, the attacker chooses a set of $2t + 1$ servers, $L$, in which all corrupted servers are included. Then, $\widetilde{H} = \{1, \ldots, n\} \setminus C$ is the set of honest servers, $H = L \cap \widetilde{H} = \{h_1, \ldots, h_{t+1}\}$ is the set of the honest servers that join the request-response process. The attacker can obtain all information in the corrupted servers and all information what those servers can observe in all phases. Furthermore, the attacker and the corrupted servers $C$ can generate messages in arbitrary way and send them to honest servers $\widetilde{H}$. Without loss of generality, we assume $c_1 < \cdots < c_t$ and $h_1 < \cdots < h_{t+1}$.

First, we list up all information the attacker views. In **Registration phase**, the attacker obtains the shares of $P$ and $D$ generated by the data owner using random polynomials $f_P$ and $f_D$ of degrees at most $t$ and $2t$, respectively. In each **Pre-computation and communication phase**, the attacker obtains the shares of a random number $R_h$ and "0" generated by honest server $h$ using random polynomials $f_{R_h}$ and $f_{0_h}$ of degrees at most $t$ and $2t$, respectively. In each **Data reconstruction phase (3-4)**, the attacker obtains $F_h$ as response which is computed by honest server $h$ according to the scheme procedure. Thus, the information the attacker views through this attack is as follows:

$$V_1 = \{f_P(c) \mid c \in C\},$$
$$V_2 = \{f_D(c) \mid c \in C\},$$
$$V_3 = \{f_{R_h}(c) \mid h \in H, c \in C\}, \; V'_3 = \{f_{R_h}(c) \mid h \in \widetilde{H} \setminus H, c \in C\},$$
$$V_4 = \{f_{0_h}(c) \mid h \in H, c \in C\}, \; V'_4 = \{f_{0_h}(c) \mid h \in \widetilde{H} \setminus H, c \in C\},$$
$$V_5 = \{F_h \mid h \in H\}.$$

Clearly, $V_1, V_2$ themselves do not leak any information about password $P$ and secret data $D$. Also, $V'_3, V'_4$ give no information to the attacker, because they are not used in **Data reconstruction phase**. On the other hand, $V_3$ and $V_4$ have the possibility to leak some additional information because they are related to the response $V_5$ through $f_{R_h}$ and $f_{0_h}$. So, we consider $(V_3, V_4, V_5)$.

Let $R_c^{(h)}$ and $Z_c^{(h)}$ be the values sent from corrupted server $c$ to honest server $h$ as a share of random number and "0", respectively, in **Pre-computation and communication phase**. Let $P'^{(h)}$ be the value that the attacker, at the beginning of **Data reconstruction phase,** sends to honest server $h$ as a request. Note that these values may not be determined from polynomials, but there exists a unique polynomial $f_{P'}(x)$ of degree $t$ which satisfies $f_{P'}(h) = P'^{(h)}$ for all $h \in H$. Hereafter, we show that $V_3, V_4, V_5$ has no additional information, unless $P \neq f_{P'}(0)$ holds.

Each $F_h$ in $V_5$ is computed by honest server $h$ as follows:

$$F_h = \left(f_P(h) - P'^{(h)}\right)\left(\sum_{h' \in H} f_{R_{h'}}(h) + \sum_{c \in C} R_c^{(h)}\right) + \sum_{h' \in H} f_{0_{h'}}(h) + \sum_{c \in C} Z_c^{(h)} + f_D(h). \quad \text{(A.1)}$$

Defining

$$\Delta_h = f_P(h) - P'^{(h)} = f_P(h) - f_{P'}(h),$$

$$f_R(x) = \sum_{h \in H} f_{R_h}(x) = \sum_{i=0}^{t} \rho_i x^i,$$

$$f_Z(x) = \sum_{h \in H} f_{0_h}(x) = \sum_{i=1}^{2t} z_i x^i,$$

Eq. (A.1) is written as

$$F_h = \Delta_h \left( f_R(h) + \sum_{c \in C} R_c^{(h)} \right) + f_Z(h) + \sum_{c \in C} Z_c^{(h)} + f_D(h).$$

Then, all values in $V_3, V_4, V_5$ are represented by the following linear equation:

$$\begin{pmatrix} \sum_{h \in H} f_{R_h}(c_1) \\ \vdots \\ \sum_{h \in H} f_{R_h}(c_t) \\ F_{h_1} \\ \vdots \\ F_{h_{t+1}} \\ \sum_{h \in H} f_{0_h}(c_1) \\ \vdots \\ \sum_{h \in H} f_{0_h}(c_t) \end{pmatrix} = \begin{pmatrix} 0 \\ \vdots \\ 0 \\ \Delta_{h_1} \sum_{c \in C} R_c^{(h_1)} + \sum_{c \in C} Z_c^{(h_1)} + f_D(h_1) \\ \vdots \\ \Delta_{h_{t+1}} \sum_{c \in C} R_c^{(h_{t+1})} + \sum_{c \in C} Z_c^{(h_{t+1})} + f_D(h_{t+1}) \\ 0 \\ \vdots \\ 0 \end{pmatrix} + M \begin{pmatrix} \rho_0 \\ \vdots \\ \rho_t \\ z_1 \\ \vdots \\ z_{2t} \end{pmatrix}, \quad (A.2)$$

where

$$M = \begin{pmatrix} A & 0 \\ E & K \\ 0 & B \end{pmatrix},$$

and

$$A = \begin{pmatrix} 1 & c_1 & c_1^2 & \cdots & c_1^t \\ \vdots & \vdots & \vdots & \ddots & \vdots \\ 1 & c_t & c_t^2 & \cdots & c_t^t \end{pmatrix}, B = \begin{pmatrix} c_1 & c_1^2 & \cdots & c_1^{2t} \\ \vdots & \vdots & \ddots & \vdots \\ c_t & c_t^2 & \cdots & c_t^{2t} \end{pmatrix},$$

$$E = \begin{pmatrix} \Delta_{h_1} & \Delta_{h_1} h_1 & \Delta_{h_1} h_1^2 & \cdots & \Delta_{h_1} h_1^t \\ \vdots & \vdots & \vdots & \ddots & \vdots \\ \Delta_{h_{t+1}} & \Delta_{h_{t+1}} h_{t+1} & \Delta_{h_{t+1}} h_{t+1}^2 & \cdots & \Delta_{h_{t+1}} h_{t+1}^t \end{pmatrix}, K = \begin{pmatrix} h_1 & h_1^2 & \cdots & h_1^{2t} \\ \vdots & \vdots & \ddots & \vdots \\ h_{t+1} & h_{t+1}^2 & \cdots & h_{t+1}^{2t} \end{pmatrix}.$$

Because all values $(\rho_0, \ldots, \rho_t, z_1, \ldots, z_{2t})$ in the last term of Eq. (A.2) are chosen by the honest servers, they have uniform distribution. Thus, if the matrix $M$ is nonsingular, then $(V_3, V_4, V_5)$ as a whole has uniform distribution and is independent from $\Delta_h, f_D, R_c^{(h)}, Z_c^{(h)}$, and the choice of $C$. The non-singularity of $M$ is evaluated as follows.

Let $E_i$ be the $i$-th row of the matrix $E$ and $K_i'$ be the matrix obtained by removing the $i$-th row from the matrix $K$. Then

$$\det M = \sum_{i=1}^{t+1} \left( (-1)^{i-1} \det \begin{bmatrix} A \\ E_i \end{bmatrix} \cdot \det \begin{bmatrix} K_i' \\ B \end{bmatrix} \right),$$

$$\det\begin{bmatrix}A\\E_i\end{bmatrix} = \Delta_{h_i} \det\begin{bmatrix}1 & h_i & \overset{A}{\cdots} & h_i^t\end{bmatrix} = (-1)^t \Delta_{h_i} \left(\prod_{c\in C}(c-h_i)\right)\left(\prod_{\substack{c'>c\\c,c'\in C}}(c'-c)\right),$$

and

$$\det\begin{bmatrix}K_i'\\B\end{bmatrix} = \left(\prod_{h\in H\setminus\{h_i\}} h\right)\left(\prod_{c\in C} c\right)\left(\prod_{h\in H\setminus\{h_i\}}\left(\prod_{c\in C}(c-h)\right)\right)$$

$$\times \left(\prod_{\substack{c'>c\\c,c'\in C}}(c'-c)\right)\left(\prod_{\substack{h'>h\\h,h'\in H\setminus\{h_i\}}}(h'-h)\right).$$

By noting

$$f_P(0) - f_{P'}(0) = \sum_{h_i\in H} \Delta_{h_i} \left(\prod_{h\in H\setminus\{h_i\}} h\right)\left(\prod_{h\in H\setminus\{h_i\}}(h-h_i)\right)^{-1},$$

we obtain

$$\det M = (-1)^t \left(\prod_{c\in C} c\right)\left(\prod_{\substack{c'>c\\c,c'\in C}}(c'-c)\right)^2 \left(\prod_{h\in H}\left(\prod_{c\in C}(c-h)\right)\right)$$

$$\times \left(\prod_{\substack{h'>h\\h,h'\in H}}(h'-h)\right) (f_P(0) - f_{P'}(0)).$$

Hence, $\det M \neq 0$ holds if and only if $f_P(0) = P \neq f_{P'}(0)$. Note that polynomial $f_{P'}(x)$ is determined by the values $f_{P'}(h) = P'^{(h)}$ sent from the attacker to the honest servers as password's shares. So, $f_{P'}(0)$ can be considered as a "guessed" password $P'$. Consequently, $(V_3, V_4, V_5)$ has no additional information if the guessed password $P'$ is not equal to the registered password $P$. This means that, no matter how the attacker acts co-operating with the corrupted servers, it cannot get additional information beyond the on-line dictionary attack.

In this attack model, the attacker is assumed to corrupt the $t$ servers in $L$. If one thinks this assumption not realistic, we can slightly modify the scheme as follows. For every possible set $L$, **Pre-computation and communication phase** is performed by the servers in $L$ (rather than all servers). When $L$ is specified by the data owner in **Data reconstruction phase**, servers use shares of random numbers and "0" for the specified $L$. The used shares must be discarded, but shares for other set $L'(\neq L)$ need not to be discarded. The security of this modified protocol is proven in a similar way to the above proof. Note that if the number of servers $n$ is $2t+1$, we need not modify the scheme.